\documentclass[a4paper,UKenglish,cleveref, autoref, thm-restate]{lipics-v2019}
\usepackage[nolist]{acronym}
\nolinenumbers

\lstdefinelanguage{NuSMV}{
  keywords={MODULE, VAR, DEFINE, ASSIGN, init, next, TRUE, FALSE, case, esac, SPEC, LTLSPEC, G},
  keywordstyle=\bfseries,
  ndkeywords={class, export, boolean, throw, implements, import, this},
  ndkeywordstyle=\color{darkgray}\bfseries,
  identifierstyle=\color{black},
  sensitive=false,
  comment=[l]{--},
  morecomment=[s]{/*}{*/},
  commentstyle=\color{purple}\ttfamily,
  stringstyle=\color{red}\ttfamily,
  morestring=[b]',
  morestring=[b]"
}

\newcommand{\sysname}[0]{\emph{SLICED}}

\title{Compositional Formal Analysis Based on Conventional Engineering Models}

\author{Tyler D. Smith}{Adventium Labs, United States \and \url{https://www.adventiumlabs.com} }{tyler.smith@adventiumlabs.com}{https://orcid.org/0000-0002-8460-6573}{}

\author{Ryan Peroutka}{Adventium Labs, United States \and \url{https://www.AdventiumLabs.com} }{ryan.peroutka@adventiumlabs.com}{https://orcid.org/0000-0003-0906-8776}{}

\author{Dr.~Robert Edman}{Carnegie Mellon University, United States \and \url{https://robertedman.com/} }{redman@andrew.cmu.edu}{} {} % TODO ORCID

\authorrunning{T. Smith, R. Peroutka, R. Edman}

\Copyright{Adventium Labs} %TODO mandatory, please use full first names. LIPIcs license is "CC-BY";  http://creativecommons.org/licenses/by/3.0/


\category{} %optional, e.g. invited paper

\relatedversion{} %optional, e.g. full version hosted on arXiv, HAL, or other respository/website
\supplement{}%optional, e.g. related research data, source code, ... hosted on a repository like zenodo, figshare, GitHub, ...

\funding{This material is based upon work supported by the U.S. Naval Air Warfare Center, Aircraft Division under contract no. N68335-16-C-0506
and NASA Marshall Space Flight Development Center under contract No. 80NSSC19C0075.
Any opinions, findings, and conclusions or recommendations expressed in this material are those of the author(s)
and do not necessarily reflect the views of the U.S. Naval Air Warfare Center or NASA Marshall Space Flight Development Center.
NAVAIR Public Release 2020-113. Distribution Statement A - ``Approved for public release; distribution is unlimited.''}

\usepackage{fancyhdr}
\EventShortTitle{preprint}
\ArticleNo{1}
\begin{document}

\maketitle

\begin{abstract}
Applications of formal methods for state space exploration have been successfully applied to evaluate robust critical software systems.
Formal methods enable discovery of error conditions that conventional testing may miss, and can aid in planning complex system operations.
However, broad application of formal methods has been hampered by the effort required to generate formal specifications for real systems.
In this paper we present State Linked Interface Compliance Engine for Data (SLICED),
a methodology that addresses the complexity of formal state machine specification generation by
leveraging conventional engineering models to derive compositional formal state models and to generate formal assertions on the state machines.
We demonstrate SLICED using the Virtual ADAPT model published by NASA and validate our results by replicating them using Simulink.
\end{abstract}

\section{Introduction}

Formal methods for system verification, particularly via \emph{model checking} have been in use since the 1980s.
A common use of model checking is to create a representation of the system under analysis as a \emph{\ac{FSM}} with assertions
to be verified written in \emph{temporal logic}. A \emph{model checker} is then used to evaluate the \ac{FSM} against the temporal
logic assertions, providing counterexamples for any assertions for which an exception exists. Several strategies have been developed
for representing and exploring finite state machines, notably \emph{symbolic model checking}, which represents sets of
states as Boolean functions. Exploration of these functions can be accomplished via \emph{\ac{BDD}} manipulation, which treats the
state model exploration as a graph traversal, or by \emph{\ac{BMC}}, which treats bounded length executions of the state model as a
SATISFIABILITY (SAT) problems \cite{biere2003bounded}.

A detailed survey of the practice of model checking is beyond the scope of this paper (See \cite{biere2003bounded} for an overview).
However, a theme in model checking methodologies encountered by the authors
is the need to translate or adapt one's system design to an \ac{FSM} with temporal logic assertions.
Many such system designs are already written as conventional engineering models (e.g., AADL, SysML, Simulink).
\footnote{For the purpose of this paper, \textbf{conventional engineering models} are structured models that follow a well-defined semantic structure and describe
stateful system components, but that do not explicitly provide \acp{FSM} or only provide \acp{FSM} specific components.}
The SLICED methodology outlined in this paper provides a framework for
generation of an \ac{FSM} and associated assertions from conventional engineering models, as well as recommendations for
addressing scalability concerns associated with the analysis of large engineering models.

\paragraph*{Background}
SLICED originated as an analysis methology for generating \ac{FSM} and temporal logic assertions from \ac{AADL}.
\ac{AADL} is a language for describing embedded software systems in semantically precise terms \cite{AS5506C}.
SLICED has subsequently evolved to also include generation of \acp{FSM} and temporal logic assertions from SysML
and Simulink.  SLICED relies on well-defined component \emph{archetypes} to generate
standardized \emph{behavior} \ac{FSM} for individual system components that can be assembled into a composite \ac{FSM} (see \autoref{behavior-definition}).

\begin{definition}\label{behavior-definition}
  A component's \textbf{significant states} are states that affect or are affected by other components.
  In the context of SLICED, a component's \textbf{behavior} is a specification of its significant states,
  how external events affect its active significant state, and how its significant states or transitions between significant states affect other components.
\end{definition}

\ac{ADAPT} is a testbed for evaluating electrical systems and injecting faults in a controlled manner. \ac{ADAPT}
consists of a power supply, batteries, sensors, and actuators. \ac{ADAPT} has several hundred components \cite{mengshoel2008diagnosing}.
NASA created Virtual ADAPT to provide a digital equivalent of \ac{ADAPT}. Virtual ADAPT is a Simulink model
of the majority of \ac{ADAPT} \cite{VirtualAdapt}.

We exercise the SLICED methodology using the symbolic model checker NuSMV. NuSMV provides
the ability to specify \emph{Modules}, which are collections of variables and transitions analogous
to subcomponents in a system design \cite{cimatti1999nusmv}.

\paragraph*{Scope}
This paper describes the SLICED methodology and how it applies to the Virtual ADAPT source model.
We provide examples of the generated \ac{FSM} as specified in the \ac{SMV} language. We provide performance
metrics generated from running the NuSMV solver on our generated \ac{FSM}. We conclude with recommendations for
future research.

\section{Motivation}
\label{sec:motivation}

\paragraph*{Software Integration Cost}

\begin{figure}[t]
\centering{\includegraphics[width=1\linewidth]{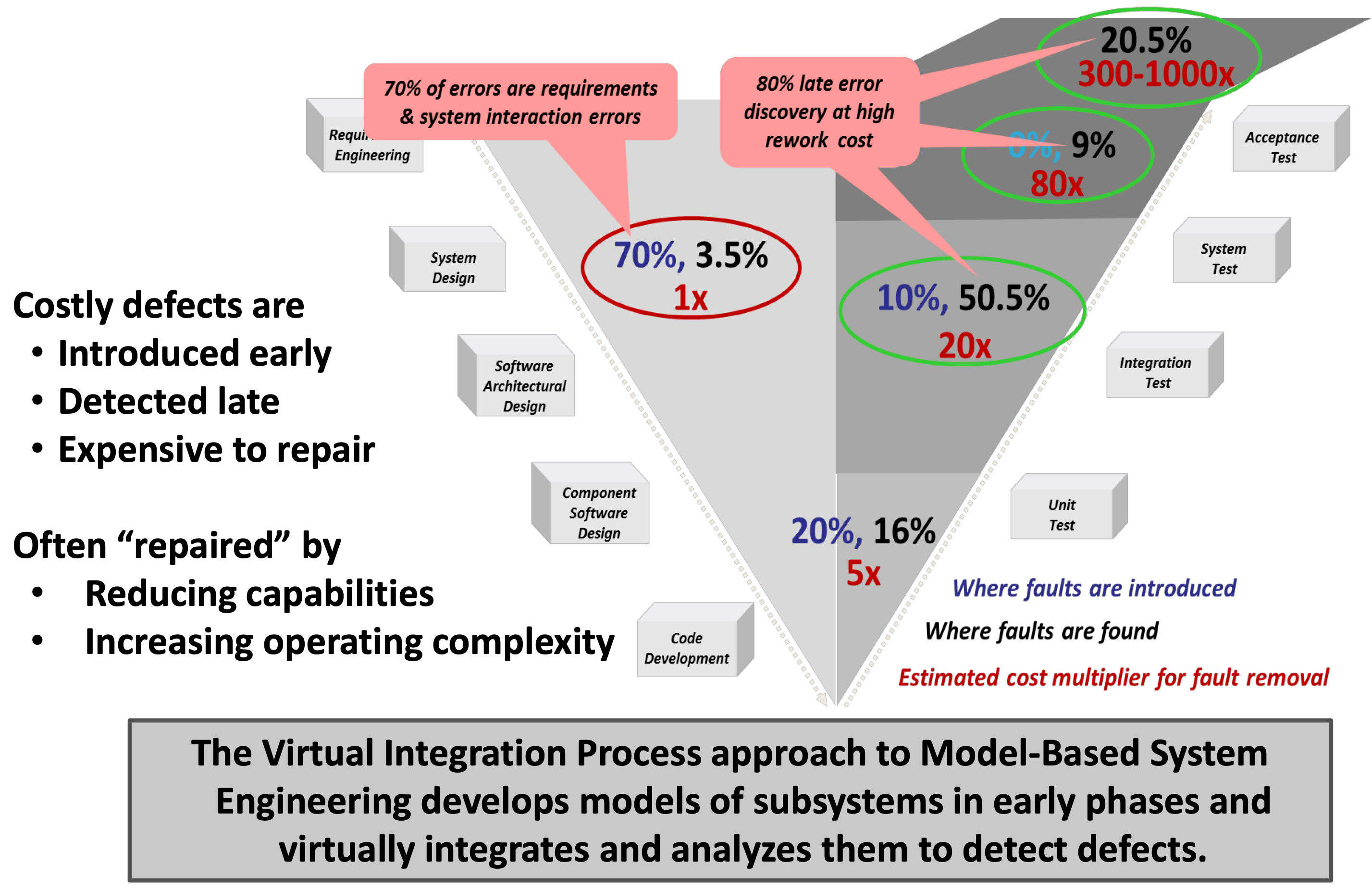}}
\caption{Costs of incompatibilities over the lifetime of a project \cite{planning2002economic} \cite{galin2004software} \cite{barry1981software}.}
\label{fig:virtual-integration}
\end{figure}

Software integration is a major cost driver in avionics development.
Integration problems often go undetected until late in the development process, when the impact to cost and schedule to fix such issues
is much higher (as shown in \autoref{fig:virtual-integration}).

\paragraph*{Limitations of Current Approaches}
A variety of efforts in recent years have leveraged \ac{MBSE} to reduce the risk of software integration, such as \ac{FACE} \cite{FACE30}.
These efforts have garnered industry momentum toward modeling as a key element of systems engineering.
However, system faults associated with the underspecified \emph{behavior} of integrated components remain problematic.
For example, in 1999 the Mars Polar Lander is believed to have crashed because of unspecified behavior of Hall Effect sensors on its landing legs \cite{spacecraft-accidents}.
The behavior was known to the landing leg engineers, but was not communicated to the software engineers.
Integrated system behavior can also manifest in more subtle errors, such as timing errors in software integration due to inconsistent communication
paradigm assumptions in software components found in a recent U.S. Army Study \cite{JCAShadowIntegration15-findings}.

Formal modeling of component behavior has potential to address these issues. Formal modeling of the Mars Polar Lander landing legs
could have communicated the legs behavior and enabled automated detection of integration errors in the interaction between the landing legs and software.
Formal analysis can uncover unlikely system configurations that may be difficult to evaluate using conventional testing methods.
However, there are two significant factors that complicate application of formal methods that must be addressed
before formal methods can be viable in avionics systems engineering workflows. First, formal methods
are outside of the expertise of many systems engineers. Although training can bridge this gap, tools and
methodologies that reduce the effort required to employ formal methods could provide a low-cost-high-return
alternative to extensive training. Second, current applications of formal methods analysis are not geared toward \emph{integration}
of components from different organizations or teams.

\paragraph*{Contributions of this Paper}

\begin{figure}[t]
\centering{\includegraphics[width=1\linewidth]{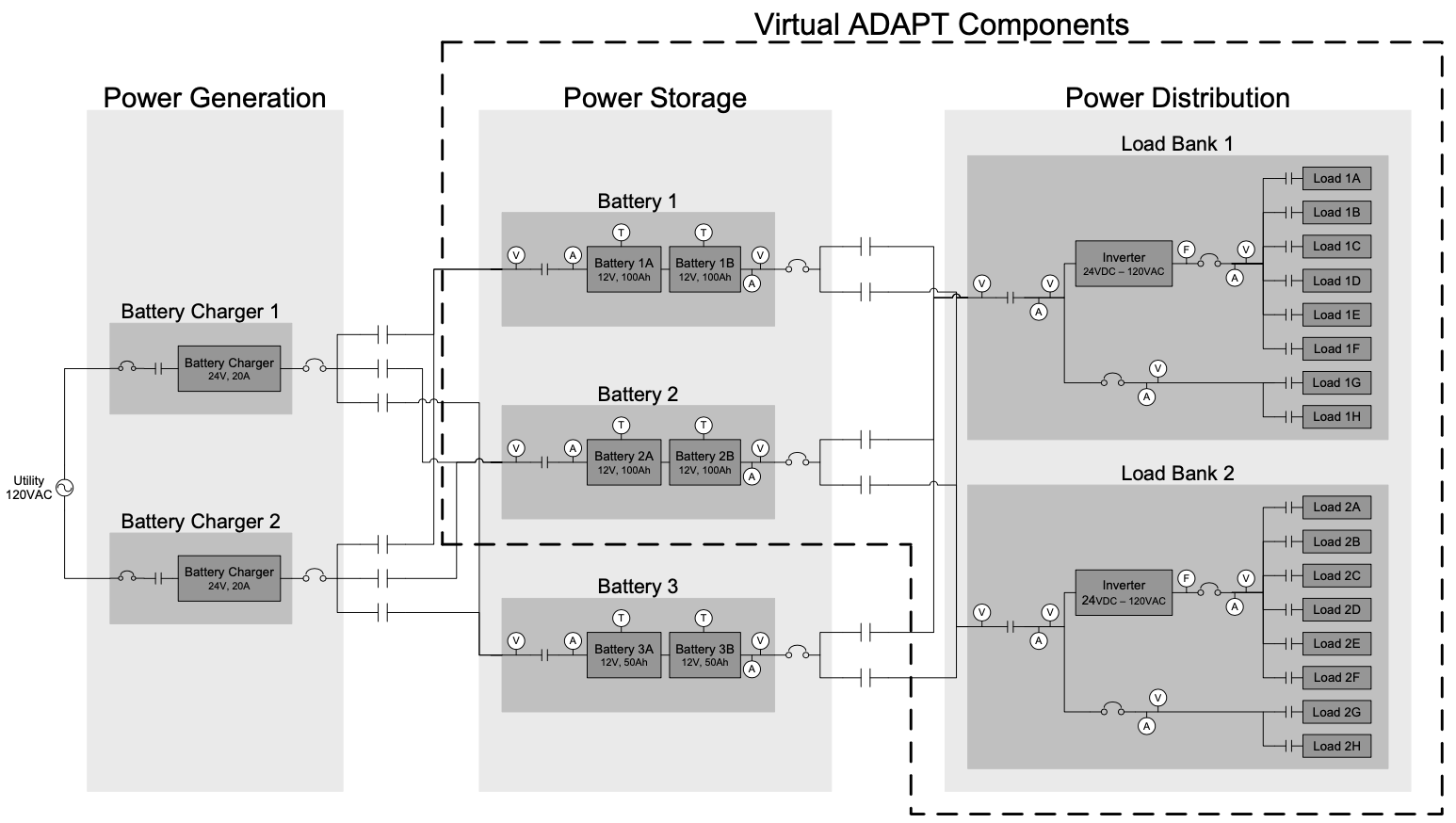}}
\caption{ADAPT components, with components modeled in Virtual ADAPT highlighted.}
\label{fig:nasa-virtual-adapt}
\end{figure}

The objective of this SLICED project is to make formal methods capabilities accessible to systems engineers by leveraging
existing formal methods tools, \ac{MBSE} environments, and models.
This paper describes how SLICED addresses two core limitations of the state of art using a \emph{compositional}
approach to formulating formal methods specifications in terms of standardized \emph{component architetypes}
and a methodology for generating formal specifications and assertions from conventional engineering models.
This paper also explores methods for creating formal assertions to check both for potential error conditions and to
generate plans for error recovery.
This paper describes a case study demonstrating a novel approach translating a Simulink model to modularized NuSMV state machine specifications.
This paper uses the NASA Virtual ADAPT model for evaluation.
The overall architecture for \ac{ADAPT} and the elements implemented in Virtual ADAPT are shown in \autoref{fig:nasa-virtual-adapt} from \cite{VirtualAdapt}.

Our approach in SLICED is based on \ac{ACVIP}. \ac{ACVIP} provides early detection of integration errors through model integration \cite{acvip}.
Our first automated implementation of the SLICED methodology was based on \ac{AADL},
which provides a lexicon for clearly and consistently describing embedded software components \cite{AS5506C}.
The most common language for \ac{ACVIP} modeling is \ac{AADL}, which is an embedded computing systems modeling language.
Although the approach described in this paper does not use \ac{AADL} directly (ADAPT is provided by NASA in Simulink, not AADL), the component prototypes used to
establish interface boundaries in ADAPT (e.g., battery and relay) are defined with intent and granularity informed by types in \ac{AADL} (e.g., thread and bus).
Simulink models provide a test platform for simulation-based evaluation. SLICED takes a Simulink model,
abstracts the inner complexity of its components, creates an \ac{FSM}, and evaluates that \ac{FSM} to generate scenarios for evaluation via simulation.

The notation in this paper follows that used by Biere et al. who describe an \ac{FSM} as a Kripke structure \(M = (S,I,T,L)\) and

\begin{itemize}
  \item S is a set of states.
  \item I is a set of initial states.
  \item T is a set of transitions between states.
  \item L is a labeling of states with atomic propositions that hold in each state.
\end{itemize}

A path \(\pi\) in this structure is a sequence of states representing on execution of the system \cite{biere2003bounded}.

\paragraph*{Research Hypotheses}

There is an inherent loss of fidelity that occurs when tranlating from a highly detailed Simulink model to a abstract behavior model.
This loss is acceptable because the objective of SLICED is not to perform a precise simulation of the system (Simulink already does that); rather, the
objective of SLICED is to \emph{generate} scenarios that can be evaluated using precise simulation. \autoref{sliced-abstraction-claim} states this notion formally.
There are error conditions that SLICED may not find, but those it does find are replicatable in Simulink.
\footnote{
``\emph{Error}'' as used in \autoref{sliced-abstraction-claim} and throughout this paper refers to a deviation from a desired system state \cite{avizienis2004basic}.
}

\begin{claim}\label{sliced-abstraction-claim}
  Let P be a source model in some language that can be executed or simulated.
  Let R be the set of system states possible in P and r in R be a single system state.
  Let M be a Kripke model decomposed from P.
  A \emph{faithful} decomposed of P to M will yield a set
  of states S in which each state s maps to one or more states r in R.
  A path \(\pi\) on M is called replicable if there exists a sequence of states \(\Phi\) in R
  for which each state \(s_n\) maps to one or more states \(r_{\mathrm{j...k}}\)
  and \(s_{\mathrm{n+1}}\) maps to one or more states \(r_{\mathrm{j+1...k}}\).
  An error path \(\pi\) found in a \ac{FSM} generated by decomposition from M is replicable via simulation or execution of P.
\end{claim}

%\begin{itemize}
%	\item The compositional approach of SLICED will be sufficient for performance on a large model.
%	\item Errors found in a state model approximation are relevant to a physical model.
%\end{itemize}

\section{Related Work}
\label{sec:related-work}

\paragraph*{Related Work on Component Interoperability}
The need to modularize software and hardware development to foster interoperability has driven a variety of research and standardization efforts.
Notably, the \ac{FACE} Technical Standard standardizes the concept of a Unit of Portability, a software component whose interface
is defined using a standardized modeling language \cite{FACE30}. The well-defined boundary of \ac{FACE} Units of Portability enables
compositional analysis earlier than would otherwise be possible \cite{interorgintegration}.
Similarly, the \ac{HOST} standard defines an approach for modular hardware components \cite{HOST12}.

\paragraph*{Related Work on Application of Formal Methods}
A variety of efforts have explored the potential of formal methods application to avionics engineering.
The \ac{DARPA} \ac{HACMS} project successfully generated code from a formally verified model of assume-guarantee contracts \cite{formalsecureair}.
The \ac{HACMS} project used a tool called \ac{AGREE} developed by Collins Aerospace that generates
formal, invariant-oriented specifications from \ac{AADL} models. \ac{AGREE} generates the formal specification
from the model, but is not capable of reasoning about infinite time conditions and requires that the user
explicitly state assumptions and guarantees about system conditions. Vestal describes
application of finite state automata to embedded software scheduling \cite{vestal-schedule-verification}.
Boddy et al. generated a processor schedule using a formal problem specification derived from an \ac{AADL} model \cite{spica}.
Boddy's approach treats existing properties in an \ac{AADL} model such as periodic thread deadlines as constraints
and uses them to express a scheduling problem in terms of constraints derived from the model.

\paragraph*{Related Work on Model Checking}
This paper uses NuSMV, a symbolic model cheker that supports \ac{LTL} and \ac{CTL} model checking \cite{cimatti1999nusmv}.
A variety of prior efforts use NuSMV. For example, Szpyrka et al. describe a process for generating NuSMV specifications from
Petri Nets \cite{szpyrka2014methods}. Most closely related to this project, Meenakshi et al describe a similar methodology for
translating Simulink models to NuSMV in \cite{meenakshi2006tool}.
However, the methodology described by Meenakshi et al is focused on naive NuSMV generation from a given Simulink model.

\paragraph*{Related Work on Compositional State Modeling}
Ranganath et al. described an approach to modeling communication patterns of medical devices \cite{commpatternsmedical}.
Ranganath's approach aims to reduce the compotational burden of reasoning about integrated systems by abstracting
details behind standardized interfaces. Beurdouche et al. used a compositional approach to modeling the state of \ac{TLS} communication, creating a
composite state machine by assembling state machines of individual components \cite{tls-cts}.

\paragraph*{SLICED as Associated to Related Work}
The aim of SLICED and the focus on this paper is on generation of \emph{abstract}, \emph{interoperable} specifications to allow formal analysis
of systems composed of models from multiple parties. SLICED is an extension of the interface specification approaches of \ac{FACE} and \ac{HOST}
with an aim of enabling analysis akin to that performed by Beurdouche. SLICED uses a modularization approach similar to the communication patterns
described by Ranganath, abstracting the patterns into standardized components akin to those specified by \ac{AADL}.

%Extensive research has been conducted into simulation capabilities for evaluation of component behaviors.
%For example, NASA's VirtualADAPT project (used to validate the results of this paper in \autoref{sec:validation})
%provides a framework for simulation of an avionics system to explore potential emergent behaviors.
%However, simulation cannot provide the mathematical rigor of a formal analysis.
%\footnote{https://github.com/nasa/VirtualADAPT}

%%%%%%%%%%%%%%%%%%%%%%%%%%%%%%%%%%%%%%
%%%%%%%%%%%%%%%%%%%%%%%%%%%%%%%%%%%%%%
%%%% System Model %%%%%%%%%%%%%%%%%%%%
%%%%%%%%%%%%%%%%%%%%%%%%%%%%%%%%%%%%%%
%%%%%%%%%%%%%%%%%%%%%%%%%%%%%%%%%%%%%%

\section{System Model}

SLICED takes a compositional approach to state modeling, requiring that the user provide
state machine specifications for each component that describe its behavior as manifest
at its logical boundary. Applied to a Simulink model, this means the user treats individual blocks (e.g., a battery) as
black boxes, describing their behavior only in terms of what goes in to the component and what comes out.

\begin{figure}[t]
\centering{\includegraphics[width=1\linewidth]{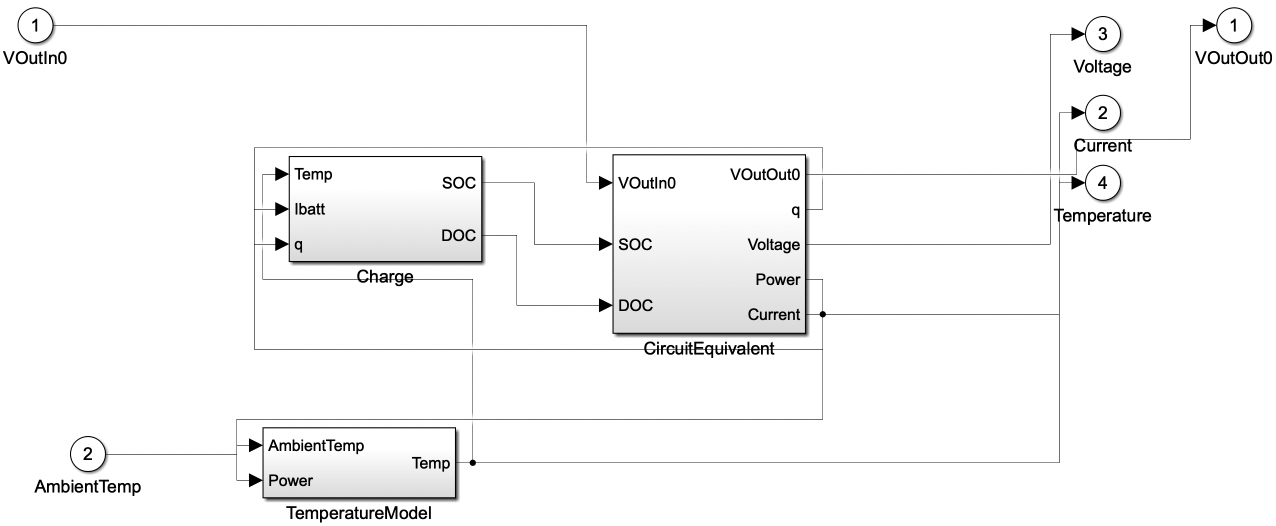}}
\caption{Battery 1 in the ADAPT model. SLICED treats the battery as a black box, tracking only the fault status of the battery and the power draw.}
\label{fig:simulink-battery1}
\end{figure}

For the ADAPT example application of SLICED, we defined five types of components, each with behavior specified at their boundary.
Doing so enabled us to abstract the inner complexity of the components, losing model fidelity but enabling formal analysis.
\autoref{lst:battery-nusmv} shows an example of a SLICED module abstracting the Simulink specification shown in \autoref{fig:simulink-battery1}.
The states listed in \autoref{lst:battery-nusmv} are taken from a secondary specification provided by NASA with the \ac{ADAPT} model.
\footnote{https://github.com/nasa/VirtualADAPT/blob/master/MATLAB/ADAPTComponents.m}

\begin{lstlisting}[language=NuSMV, label={lst:battery-nusmv}, caption={Battery module specified in NuSMV}]
MODULE Battery(output1, output2, capacity)
VAR
  state : {nominal,dead,underRepair};
DEFINE
  supplyingPower := (state = nominal);
  draw := (output1.draw + output1.draw);
ASSIGN
  init(state) := nominal;

  next(state) := case
    (draw > capacity) : dead;
    ((state = dead) & (draw = 0)) : underRepair;
    ((state = underRepair) & (draw = 0)) : nominal;
    TRUE : state;
  esac;
\end{lstlisting}

\begin{figure}[t]
\centering{\includegraphics[width=1\linewidth]{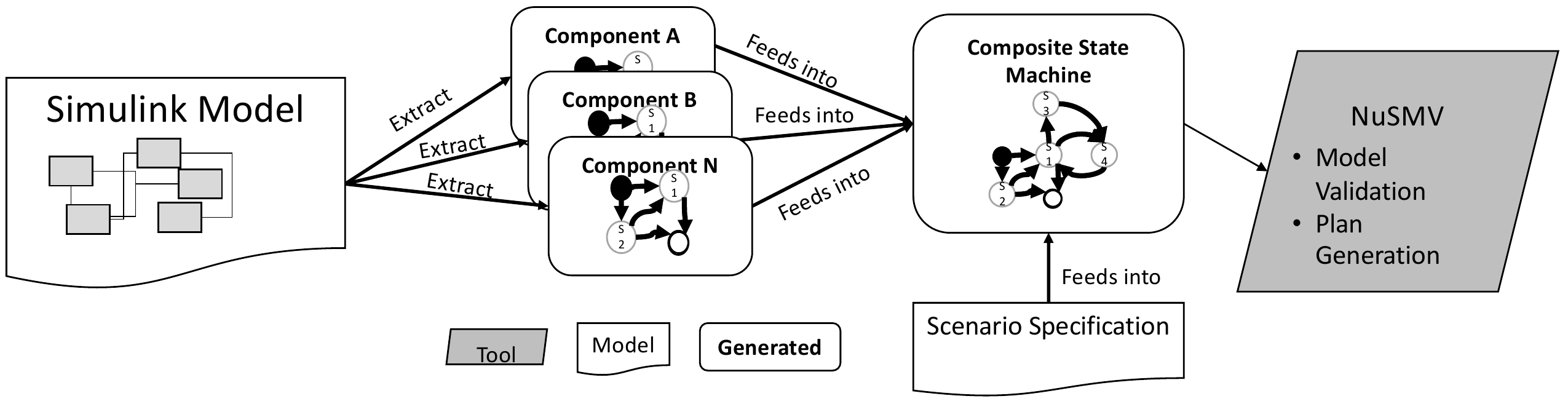}}
\caption{SLICED Architecture}
\label{fig:architecture}
\end{figure}

SLICED combines these state specifications with global system configuration (such as processor schedules, message routing, or indercomponent dependencies)
to create a \emph{composite state machine} (this data flow is shown in \autoref{fig:architecture}).

\begin{definition}\label{csm-definition}
A composite state machine is a deterministic finite automata formed from the powerset construction of independent component state machines and system configuration.
\end{definition}

By limiting the scope of each component's state machine to its behavioral interface,
SLICED can minimize the effective state space that it must analyze.
The limited scope of each component's state machine reduces the risk required to reconfigure
an integrated system design, for example by adding or removing components as shown in \autoref{sec:performance},
because the connections between components are well defined and clearly specified.
A naive deterministic finite automata construction from multiple state machines
results in the powerset construction of all of their component states, as described in \autoref{csm-definition}.
Our definition of the composite state machine is similar to the \emph{Interface Featured Time Automata} described by Cledou et al. \cite{cledou2017composing}.

A core element of the SLICED hypothesis is that discretized events in a system model can yield results that are useful for a real system. Use of a
discrete clock enables us to model timed automata (strictly, \emph{ordered} automata) where assumptions about the duration of events can be applied.
A further branch of study, the application of statistical methods to timed automata, is beyond the scope of this paper.

\paragraph*{Approaches to Finite State Machine Generation}

There are several potential levels of fidelity in translation of an electrical systems model (such as Virtual ADAPT) to a formal specification.
All of these approaches treat individual components as black boxes and construct a formal specification as an assembly of
black-box modeled components. Each level adds fidelity to the black box models, progressively them from ``black'' to ``white.''

\begin{enumerate}
	\item \textbf{State only translation:} Make no assumptions about the relationships between components. Model component states only as expressed verbatim in the source model.
	\item \textbf{State and connection translation:} Assume that error states propagate along communication paths described in the source model. Assume some source model
	states are ``good'' and some are ``bad'' and that good and bad state properties propagate (e.g., if a component is in a bad state, anything connected to it is also in a bad state).
	\item \textbf{State and connection translation with directionality:} Same as State and Connection Translation, but assume directionality in connections can be determined from the source model
	(that is, bad component state only propagates to \emph{downstream} components.) This assumes that cycles \emph{are} allowed, but that connections are directed.
%	\item \textbf{State and Connection translation with directionality and binary signals} Same as state and connection translation with directionality, but also assumes binary (on/off) values on
	%all connections. Assumes naive propagation of on/off values along lines of directionality.
	\item \textbf{State and connection translation with directionality and discrete signals:} Instead of treating connections as binary on/off,
	treat connections as capable of holding one of a fixed set of values (e.g., value from 0\dots9).
	\item \textbf{Discrete models of component behavior:} Model the internals of each component in the model, discretizing and bounding the values but otherwise fully translating its behavior
	to a formal specification.

\end{enumerate}

\paragraph*{SLICED Pseudocode}

SLICED employs approach 4, \textbf{State and connection translation with directionality and discrete signals:}.
This is in contrast with the higher-fidelity translation presented by Meenakshi et al. \cite{meenakshi2006tool}.
The SLICED implementation is similar for most modeling languages (prior implementations have used \ac{AADL} and SysML).
The SLICED implementation for \ac{ADAPT} and Simulink is as follows:

\begin{enumerate}
	\item Iterate over all elements in the simulink model (using the library \texttt{org.conqat.lib.simulink}~\footnote{https://www.cqse.eu/en/products/simulink-library-for-java/overview/}) in a depth-first search.
	\item For each leaf block, identify the \emph{archetype} of the block (e.g., relay, battery, junction). Create a new MODULE in SMV for that block using a standard prototype.
	\begin{enumerate}
		\item The ADAPT Simulink model does not define component archtetypes, so we relied on component naming conventions to differentiate components.
    For future work an \ac{AADL} model of \ac{ADAPT} could enable stronger semantic consistency in component definitions.
	\end{enumerate}
	\item As the search ascends toward the root, continue creating MODULES for each block that has a corresponding archetype.
	\item For each \emph{line} contained in each block, add input or output parameters to the appropriate modules based on the start and end points of each line.
  This step is similar to the process of creating Reo connectors described in \cite{cledou2017composing}.
  \item For each \emph{line} that traverses multiple blocks, follow the line until a SLICED-abstracted component is found. When the SLICED-abstracted component is
  found, add it as an input or an output parameter as described in the previous step.
\end{enumerate}

\subsection{Subsystem Merging}
An additional performance improvement strategy employed by SLICED is problem reduction based on \emph{effective}
communication. In software state analysis, SLICED uses intercomponent communication configuration to limit
the size of the composite state machine by excluding composite states that are not reachable using a given communication configuration,
similar to the ``port linking'' approach described by Cledou et al. \cite{cledou2017composing}.
As with Cledou et al., we treat all signals uniformly (that is, in a hardware model a component's power draw is treated as a form of communication with the component supplying the power).
SLICED uses a stragety similar to Shannon's expansion to reduce the complexity of hardware model \acp{FSM} \cite{symbolic-model-checking-clarke}.
This paper focuses on the latter, as it is most applicable to the ADAPT model.

\subsubsection{Subsystem Merging in ADAPT}
\label{sec:component-merging-in-adapt}
In the ADAPT model there are four banks of actuators, two with two actuators and two with six actuators.
Using our actuator module definition (\autoref{lst:actuator-nusmv}), this results in 16 total actuators, each with 3 states, increasing our effective
state space by a factor of \(3^{16}\).
However, inspection of the model shows that the effective communication between the each bank of actuators and the batteries and relays
all goes through a single connection (e.g., the breaker EY166 in \autoref{fig:generated-adapt}). The result is that there are a considerable numbers
of composite states of the actuators that can be discarded \emph{a priori} by collapsing the state space of six actuators into
a single \emph{effective} actuator state, shown in \autoref{lst:merged-actuator-nusmv}, whose states are limited to the available \emph{effective} states of the subsystem.
In the case of the breaker, EY166, there are \(3^6\) possible states when the subsystem is considered naively, but
there are only 13 effective states of its interaction with other components (each actuator can draw 0, 1, or 2 units of power.
There are 6 actuators, so power consumption of 0..12 is possible (thus drawlimit is set to 12 in \autoref{lst:merged-actuator-nusmv}).

  \begin{lstlisting}[language=NuSMV, label={lst:actuator-nusmv}, caption={Actuator module specified in NuSMV}]
  MODULE Actuator(input)
  VAR
    state : {nominal,nopower,faultyResistance};
  DEFINE
    draw := case
      (!(input.supplyingPower) | (state = nopower)) : 0;
      (state = nominal) : 1;
      (state = faultyResistance) : 2;
    esac;
  ASSIGN
    init(state) := nominal;
  \end{lstlisting}

  \begin{lstlisting}[language=NuSMV, label={lst:merged-actuator-nusmv}, caption={Merged Actuator module specified in NuSMV}]
  MODULE MergedActuator(input, drawlimit)
  VAR
    draw : 0 .. drawlimit;
  \end{lstlisting}

\subsubsection{Subsystem Merging Formalism}
In terms of Shannon's expansion, we divide the system problem into two sub-problems at the boundary of the connection between the
system and the parent system.
This divide is possible because of the compositional structure of the \ac{FSM}. Subsystems that interact only at the subsystem boundary (abstracting their inner components)
must define functions that aggregate their internal state (such as the addition of power consumption described in \autoref{sec:component-merging-in-adapt}).
Dependencies between components are modeled as shared variables in the \ac{FSM}.
For example, the power draw of an actuator is represented as an integer variable that is shared with upstream relays, breakers, or power supplies.
This approach to reduced subsystem representation is similar to the \emph{\ac{COI}} feature of NuSMV,
however \ac{COI} only eliminates variables not reachable from a given assertion \cite{cavada2005nusmv}.
\ac{BMC} analysis of the state model may likely have taken advantage of this feature of the model and is an opportunity for future study.

Any resulting error trace (for example, a error when the total draw is 10) can serve as input to a follow-up sub-analysis on the decomposed
subsystem. The error trace for the top level error will give us the value on the communication channel between the top level system
and the sub-system, which we then use to generate a \emph{second} assertion, this time on the sub-system. Because the only interaction between
the sub-system and the top level system is through the aggregation component, we can be confident that none of the \(3^6\) - 13 state combinations in the
sub-system could have altered our top level assessment.

%%%%%%%%%%%%%%%%%%%%%%%%%%%%%%%%%%%%%%
%%%%%%%%%%%%%%%%%%%%%%%%%%%%%%%%%%%%%%
%%%% Assertion Generation %%%%%%%%%%%%
%%%%%%%%%%%%%%%%%%%%%%%%%%%%%%%%%%%%%%
%%%%%%%%%%%%%%%%%%%%%%%%%%%%%%%%%%%%%%

\subsection{Assertion Generation}
\label{sec:assertion-generation}
SLICED deals with two classes of assertions over the same problem space, both are classical applications of temporal logical
applied by generation from conventional engineering models.
First, \sysname{} generates \emph{error discovery} assertions, which are assertions for which a counterexample indicates a system specificaiton error.
Error discovery assertions include those asserting that the system will \emph{not} enter a bad state from a good state (safety)
or that some expected state will eventually occur (liveness).
SLICED can also generate \emph{path discovery} assertions, which are assertions that the system will \emph{not} enter a good state
from a bad state. The former is useful for detecting potential system error conditions. The latter is useful
for generating plans to restore a safe system state if a bad state has occurred.

\begin{definition}\label{error-discovery-assertion-definition}
An \emph{Error Discovery Assertion} defines a condition to be avoided, such as ``the system shall never enter an unsafe state.''
A ``solution'' found by the solver for an error discovery assertion assertion indicates an error in the design.
\end{definition}

\begin{definition}\label{path-discovery-assertion-definition}
A \emph{Path Discovery Assertion} defines a condition to be reached if possible.
A ``solution'' found by the solver for a path discovery assertion indicates a plan for return to a good state.
\end{definition}

\paragraph*{Safety and Liveness Assertion Generation}
Failure assertions include both \emph{safety} assertions (describing things that should not happen)
and \emph{liveness} assertions (describing things that should happen).
SLICED generates assertions using information provided by the source model describing the expected
states of each component. For example, when dealing with software components SLICED generates \emph{liveness}
assertions for periodic threads that must always reach a \emph{final} state.
Similarly, if the source model definition of a component describes an \emph{error} state,
SLICED generates a safety assertion that the error state will never be reached.

When evaluating timing properties, SLICED uses a cyclic clock, incrementing one tick for each step
taken in the state machine evaluation. Ticks are represented at the \ac{GCD} of timing properties
of the design (e.g., for a system which expresses performance deadlines at a granularity of one millisecond, sliced
uses one millisecond for a tick).
SLICED uses a cyclic clock because all variables in a \ac{FSM} must be finite. The maximum value of the clock
is determined by the hyperperiod of all periodic elements in the source model (e.g., if the source model has threads with periods of 100, 200, and 300ms,
SLICED will use a 600ms clock cycle).
When the source model provides \emph{deadlines} for performance, SLICED evaluates the component states against its cyclic clock.

For connections between components, SLICED treats each connection as a discrete variable or as a property of an existing variable.
For connections with capacity constraints, SLICED treats the connection as a counting semaphore and creates
assertions that restrict the number of messages it can contain according to its upper bound.

\paragraph*{Path Discovery Assertion Generation}
To generate a path discovery assertion, SLICED expresses the initial (failed) state of the system in a \ac{FSM} with its \emph{initial} state set to include one or more errors.
The \ac{FSM} differs from that used to evaluate error discovery assertions because it leaves user actions
specified as non-deterministic, in effect letting the solver take on the role of the user to manipulate the system at-will.
In the \ac{FSM} a \emph{user action} is a is a transition whose guard relies on a user input event (e.g., flipping of a switch) where the source of
that event is external to the modeled system.
SLICED then generates an assertion describing the negation of the desired state, including both the original
faulty component and of all of the other components in the system.

%%%%%%%%%%%%%%%%%%%%%%%%%%%%%%%%%%%%%%
%%%%%%%%%%%%%%%%%%%%%%%%%%%%%%%%%%%%%%
%%%% SLICED FORMALISM %%%%%%%%%%%%%%%%
%%%%%%%%%%%%%%%%%%%%%%%%%%%%%%%%%%%%%%
%%%%%%%%%%%%%%%%%%%%%%%%%%%%%%%%%%%%%%

\subsubsection{SLICED Formalism}
\label{sec:sliced-formalism}

For each component in the source model, we generate a state machine \(\psi\).
Using actions \(A\) on or between these components as specified in the source model (e.g., as connections),
we assemble a composite state machine \(\Psi\).

\paragraph*{State machine semantics}
\label{sec:semantics}
The formal semantics of a state machine in \sysname{} are a $6$-tuple
$\psi=(S,A,T \subseteq\sigma \bigoplus \psi \bigoplus A, s_0 \in \psi,
E \subseteq \psi, d \in S)$.
\[
\begin{split}
  A      &\mbox{---is a set of events, or Actions,} \\
  S &\mbox{---is a set of states for the component,} \\
  T      \subseteq \sigma \bigoplus \sigma \bigoplus  A &\mbox{---is the set
    of labeled transitions,}\\
  s_0    \in S &\mbox{---is a  designated initial state,}\\
  \prod      \subseteq \Sigma &\mbox{---is a distinguished set of error
    states. } \\
\end{split}
\]

We use a shorthand for transitions of $S$, writing $s
\overset{a}{\rightarrow} s^\prime$ for an
element $t\in T$.

\paragraph*{Composite State Machine}
\label{sec:composite}
The composite state machine is built from both the
behavioral model associated with each component and connections between them.
The composite state machine is the analytic core of \sysname{} and describes the behavior of the entire system.

Formally, the composite state machine is a timed B\"uchni automaton built by viewing
each component as running independently with the addition of a
global clock. In the case of ADAPT, which does not have specific timing constraints used by \sysname{},
\sysname{} assumes events happen in discrete time steps. Transitions between states are inter-component connections,
supporting human-in-the-loop state changes through non-determinism.
For example, user-settable relays have non-deterministic open and closed state transitions to account for user actions.
Using powerset construction, we can build a \ac{DFA}
to model the behavior of the entire system~\cite{Rabin59}.
In the context of a Simulink model, a transition is an event that changes the significant state of a component as defined by its component archetype.
 The na\"ive
powerset construction is exponentially larger than the \ac{NFA} for
any individual component.

Formally, given a collection of input $n$ behavioral state machines $\psi^1,\cdots \psi^n$,
using a common data model $A$, denoted
$\psi^i=(S^i,A,T^i \subseteq\sigma^i \bigoplus S^i \bigoplus A,
s_0^i \in S, E^i \subseteq S, d \in S)$, the composite state machine is
constructed as:
\[
\begin{split}
  \Psi &= \left( \oplus \psi_i \right) \oplus A \\
       c &= \mbox{ a clock} \\
       T &= \left\{
       \begin{split}
             (\cdots,s,\cdots) \overset{a}{\rightarrow}
             (\cdots,s^\prime,\cdots)
             & \mbox{ for } s \overset{a}{\rightarrow} s^\prime \mbox{ a
               transition of } S^i \\
             & \mbox{ and S a time slice of } S_i
       \end{split}
       \right\}\\
     s_0 &= (s_0^1,s_0^2,\ldots,s_0^n,S_0) \in \Psi \\
\end{split}
\]

%The transitions are a key innovation of \sysname{}. By limiting state transitions to
%active time slices rather than using
%the full powerset construction, the number of transitions of the composite state
%machine is reduced by an order of magnitude.

\paragraph*{State Space Size}
The full state space of
the na\"ive composite state machine for a model the scale of ADAPT is combinatorially large, making it
impractical to analyze.  By using component abstractions with simplified behavior specifications,
\sysname{} dramatically reduces the size of the relevant
state space.

We achieve further reduction in the state space size by using subsystem aggregation functions table
to collapse large subsystem state spaces for top-level system analysis, then expand them to generate detailed traces for particular conditions.

\section{Evaluation Model}

\paragraph*{Virtual ADAPT}

Virtual ADAPT is a Simulink formulation of a physical testbed representing a spacecraft's electrical power system.
It allows for the injection of faults such as malfunctioning relays and sudden spikes in electrical
resistance so that the system's response can be studied.
Users can interact with Virtual ADAPT in real time through the GUI flippable switches or programmatic alteration of most model parameters.

Simulink models are represented as \texttt{.mdl} files which define a hierarchical structure of components and the connections between them through notions of subsystems, lines (same level connections), and ports (cross-level connections). Additionally, Virtual ADAPT includes MATLAB \texttt{.m} files describing fault states and state-transition logic for relevant components.

\paragraph*{Model Statistics}

Virtual ADAPT contains 11991 blocks, including the top level block.
\footnote{We did not include blocks related to fault injection in this count. Including fault injection brings the total block count to 14705.}
However, Virtual ADAPT is a hierarchical model and 73\% of those blocks are in the lower 7 of Virtual ADAPT's 13 levels of block and sub-block containment (3197 blocks in the top 6 levels).
\footnote{The top level block for this is the block named VirtualADAPT/VirtualADAPTv1}

\section{Experimentation Evaluation}

\subsection{Experiment Overview}

The objective of our experiment was to validate our approach to reducing complex system models to computationally tractable
formal state specifications through application of a set of well-defined component archetypes.
We determined success by whether we could run formal analysis on the generated state model and
that a result found for a success or error discovery assertion in a
SLICED model would also manifest in the Virtual ADAPT Simulink model.

\subsection{Application of SLICED to ADAPT}

\paragraph*{Core ADAPT Components}
Virtual ADAPT indentifies 97 of its high-level Simulink subsytems as "ADAPT Components," many of which have specified fault conditions.
\footnote{https://github.com/nasa/VirtualADAPT/blob/master/MATLAB/ADAPTComponents.m}
These components fit into several categories:  batteries, inverters, load banks, loads, circuit breakers, relays, and sensors.

Aside from loadbanks and sensors, each component is taken to be a state machine with its states and transitions straightforwardly deduced
from each type's well-defined fault modes and inferred functionality.
Additionally, this abstraction hides the detail associated with the electrical power
transer by describing component's electric states with discrete notions of power supply and consumption.

SLICED is concerned with behavioral interfaces. We identified 52 of the 602 ``SubSystem'' blocks in the top 6 levels
of Virtual ADAPT as having modelable behavioral interfaces (we identified an additional 2 at the 7th level).
We determined the applicablity of blocks via inspection of their names and uses in ADAPT, for example by
identifying any block of type ``SubSystem'' with a name containing ``battery'' as behaving according to the Battery behavior prototype.

\begin{figure}[t]
\centering{\includegraphics[width=1\linewidth]{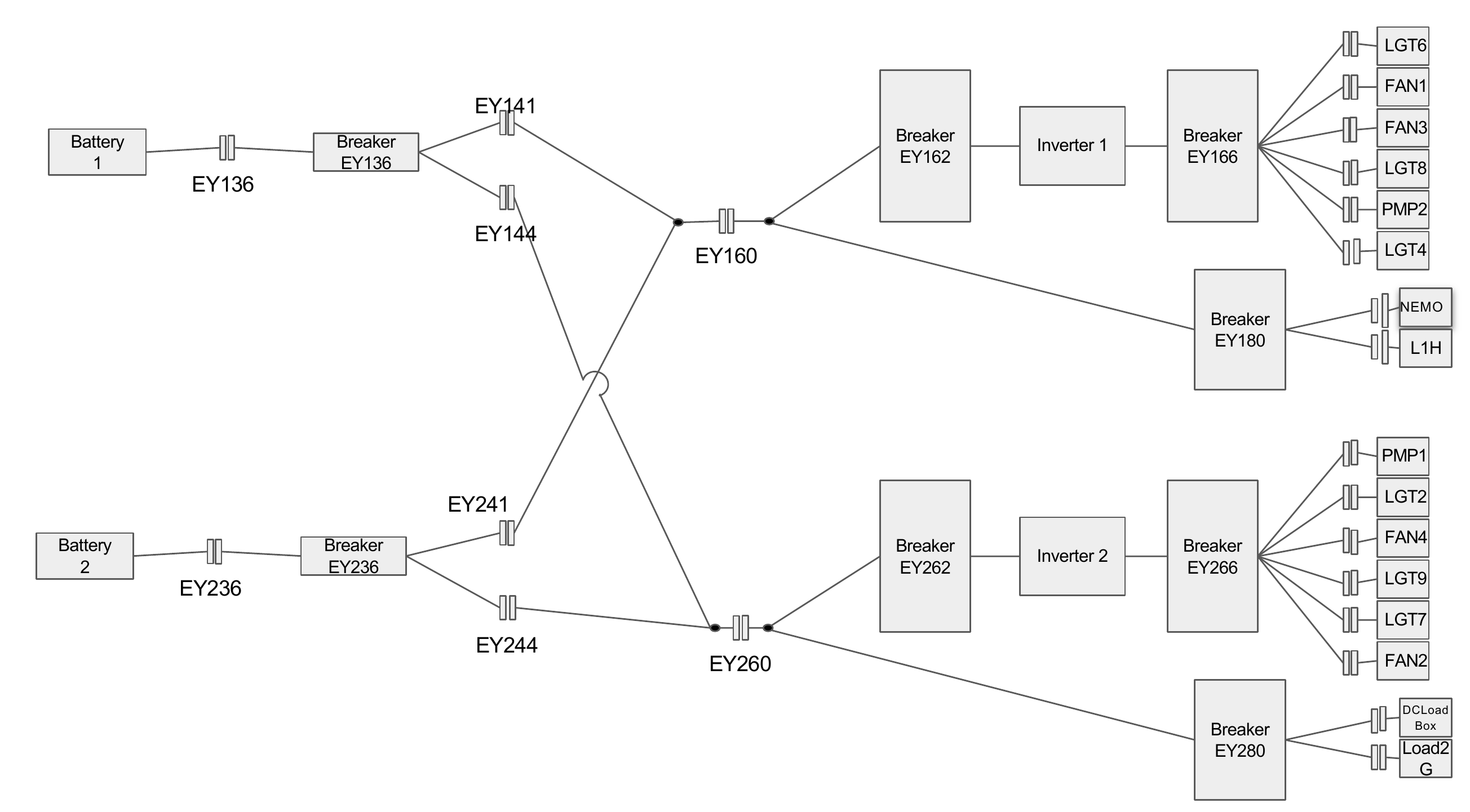}}
\caption{ADAPT Components as Reduced for Analysis by SLICED.}
\label{fig:generated-adapt}
\end{figure}

Using the 52 identified blocks, we generated a NuSMV specification including definitions of each of our re-usable component archetypes,
as well as variables defining each of the 52 blocks according to their determined archetype.
We implemented connections between the blocks using additional variables internal to each type definition.
\autoref{fig:generated-adapt} shows all of these components and their connections.

ADAPT provides connections describing both voltage and amperage on wires between components.
In creating the NuSMV representation we combined these connections into a single ``draw'' connection
representing a unitless measure of power consumption. As with our other design abstractions, this choice
results in a loss of fidelity but maintains the capacity to reason about the system in terms of relative supply and demand.

\subsection{Results}

\subsubsection{Performance Results}
\label{sec:performance}

\paragraph*{Test Platform}
%We ran our example analysis on a Windows 10 virtual machine allocated two 2.3GHz Intel i9 processor cores and 8192 MB of RAM.
We ran our example analysis on a 64 bit Windows 10 laptop with 24GB of RAM and an Intel i7 CPU running at 2.5 GHz.
We used NuSMV version 2.6.0. We measured performance of the NuSMV execution using the PowerShell \texttt{Measure-Command} cmdlet,
as shown in \autoref{lst:powershell-timing}.

\begin{lstlisting}[label={lst:powershell-timing}, caption={Powershell Invocation Example.}]
Measure-Command -Expression { nusmv .\battery_repair_full.smv > out.txt }
\end{lstlisting}

We added a simple assertion to the state model, asserting that the draw on Battery1 would be less than one.

\begin{lstlisting}[language=NuSMV, label={lst:performance-timing-assertion}, caption={Simple Assertion for Performance Timing.}]
LTLSPEC G(Battery1.draw < 1)
\end{lstlisting}

\paragraph*{Nominal Analysis}
Analysis of the generated \ac{FSM} with no performance improvements applied was computationally intractable, yielding
no results after 12 hours.

%\paragraph*{Full Model Analysis}
%We were unable to complete a analysis of the full model after 12 hours.

\begin{figure}[t]
\centering{\includegraphics[width=1\linewidth]{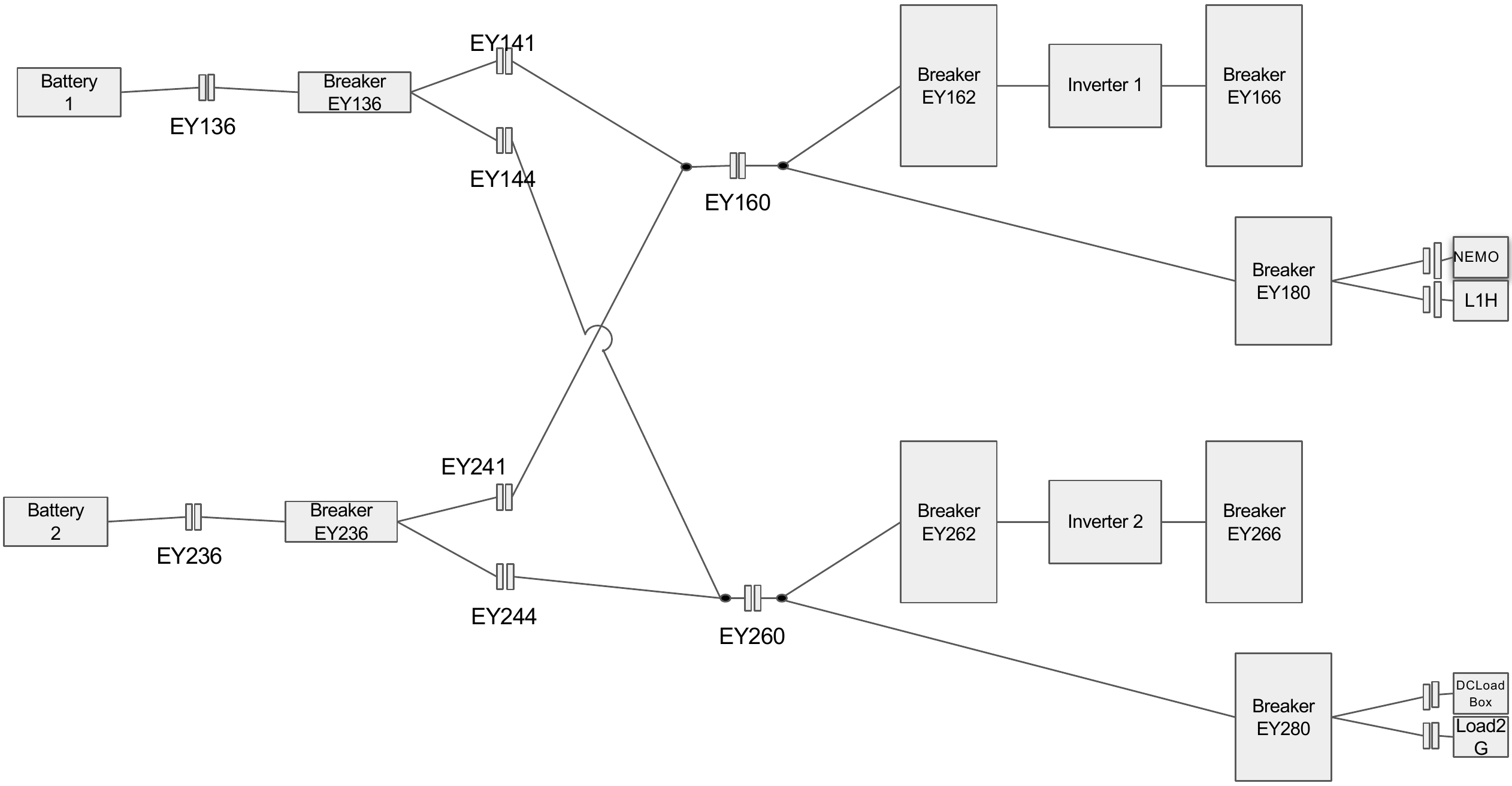}}
\caption{ADAPT Components as generated by SLICED, with 12 actuators and relays removed.}
\label{fig:adapt-reduced-one}
\end{figure}

\paragraph*{Acutator Removal}
We performed several analyses of reduced models, each time starting with the full SLICED-generated model and applying a reduction or adaptation.
First, we removed 12 actuators and their assocated relays, as
shown in \autoref{fig:adapt-reduced-one}. Analysis of this model took 3 minutes 8 seconds.

\begin{figure}[t]
\centering{\includegraphics[width=1\linewidth]{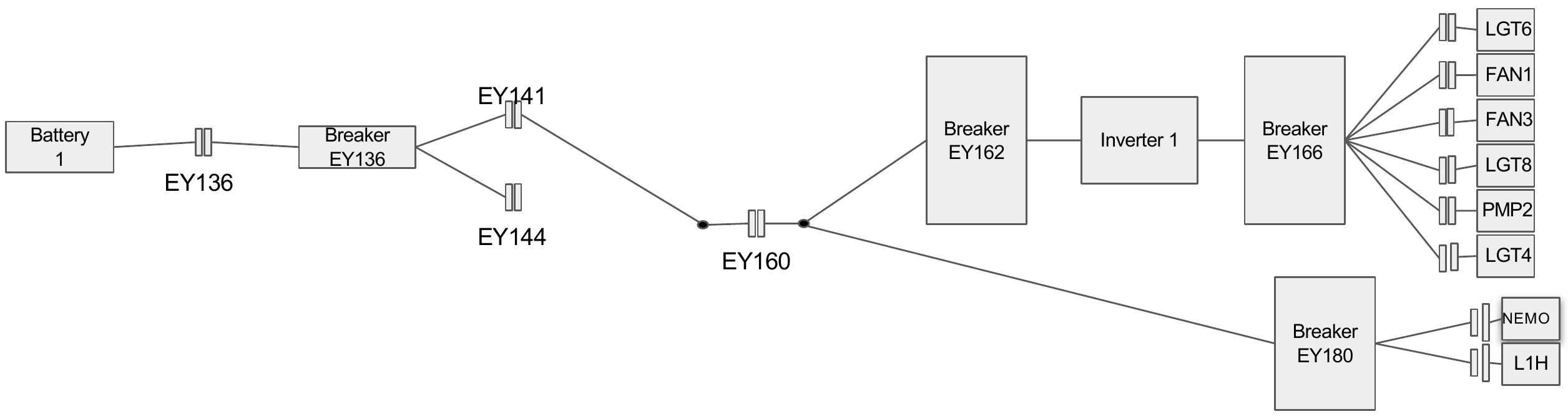}}
\caption{ADAPT Components as generated by SLICED, with the battery2 system and actuators removed.}
\label{fig:adapt-reduced-two}
\end{figure}

\paragraph*{Battery Subsystem Removal}
Second, we removed Battery2 and all of the 2-prefixed components, as shown in \autoref{fig:adapt-reduced-two}.
Analysis of this configuration took 8 minutes 34 seconds.

\begin{figure}[t]
\centering{\includegraphics[width=1\linewidth]{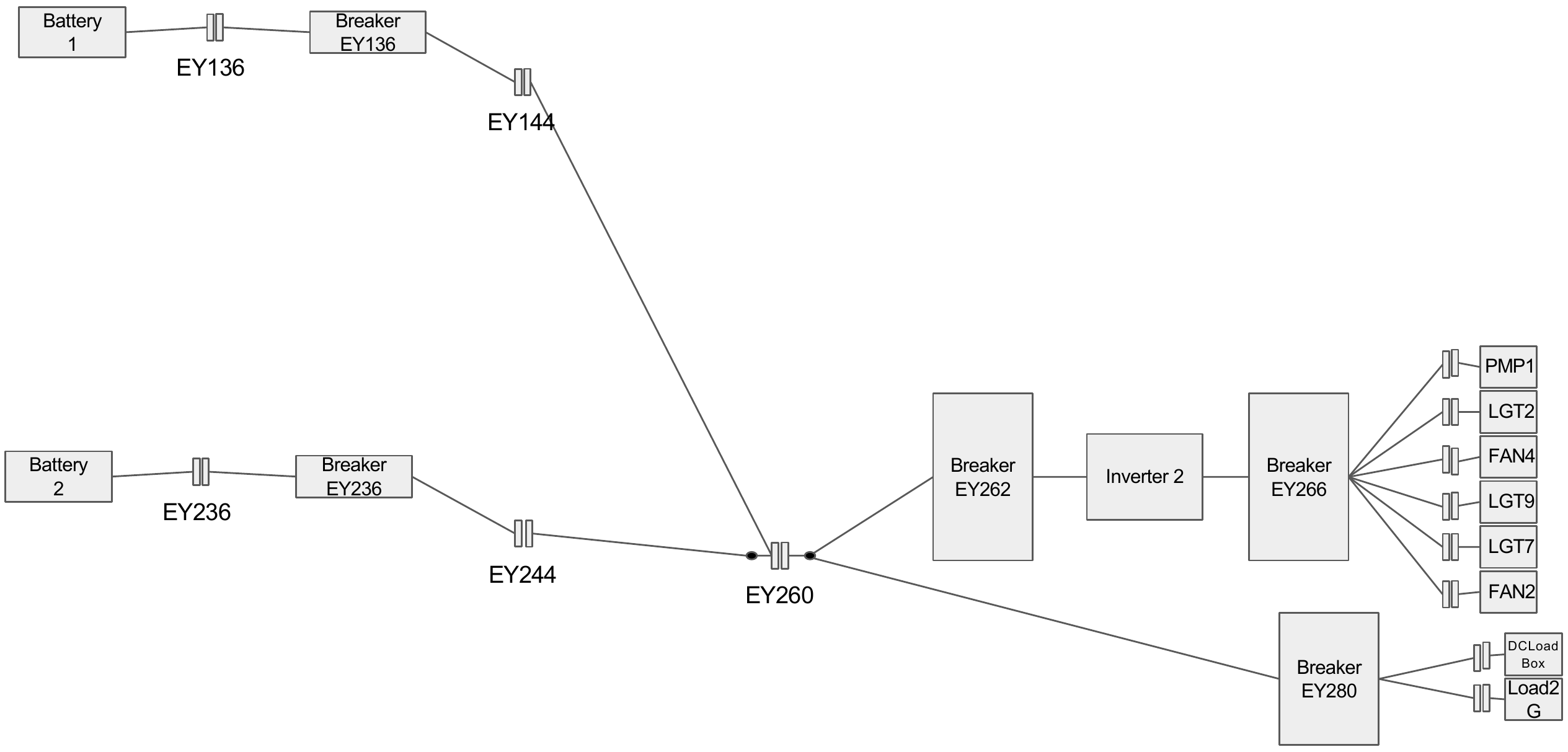}}
\caption{ADAPT Components as generated by SLICED, with the battery 1 subsystem actuators and relays removed.}
\label{fig:adapt-reduced-three}
\end{figure}

\paragraph*{Smaller Actuator Removal}
Third, we removed eight actuators and relays, as shown in \autoref{fig:adapt-reduced-three}.
Analysis of this configuration took 5 hours 25 minutes 59 seconds.

\begin{figure}[t]
\centering{\includegraphics[width=1\linewidth]{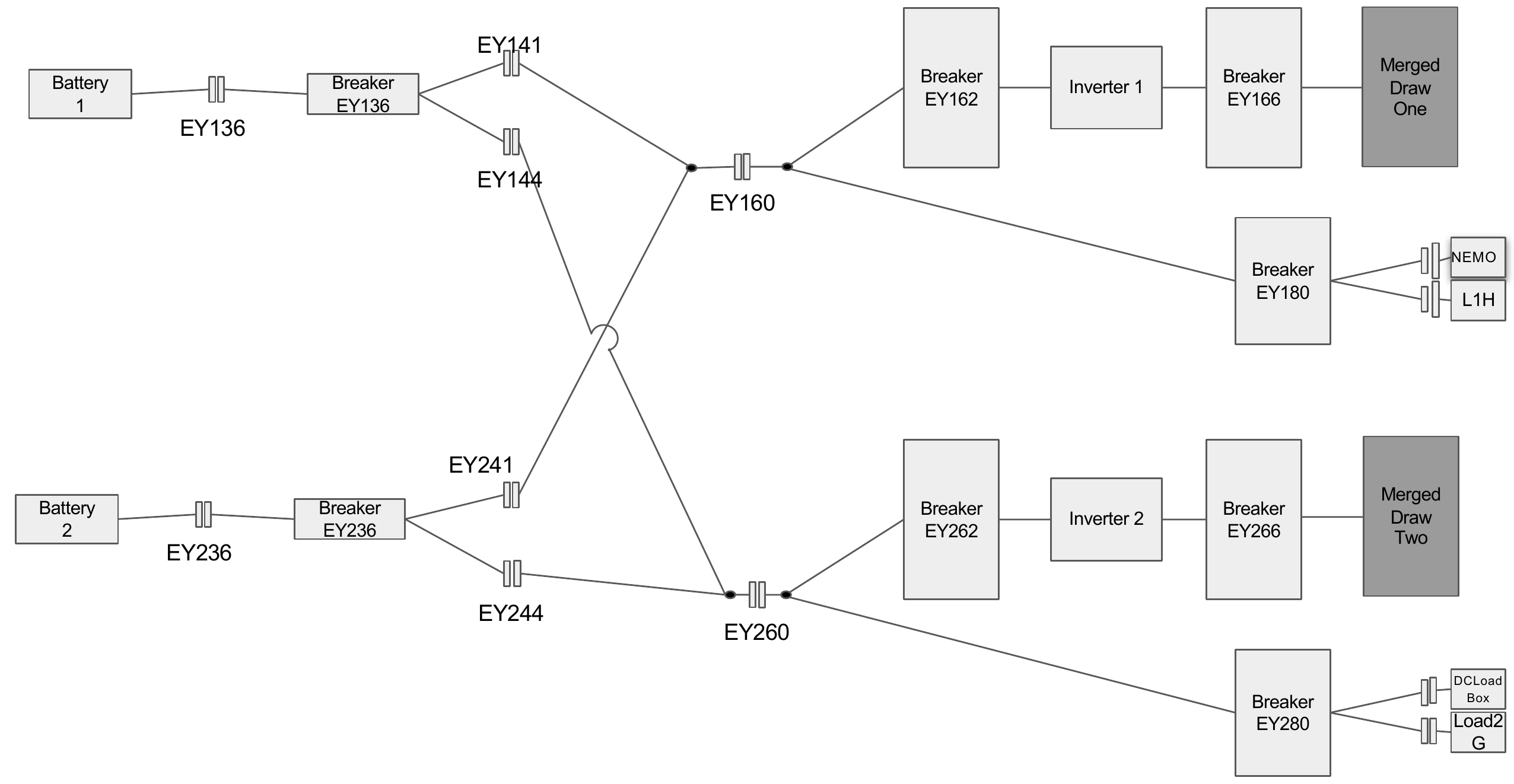}}
\caption{ADAPT Components with two banks of actuators merged into their effective representations.}
\label{fig:adapt-merged-one}
\end{figure}

\paragraph*{Actuator Subsystem Collapse}
As discussed in \autoref{sec:component-merging-in-adapt} we merged the relays and actuators in two subsystems into abstractions
representing only the possible power consumption of the subsystem (see \autoref{fig:adapt-merged-one}).
 Analysis of this configuration took 26 minutes 51 seconds.
This result is of particular note because this reduction did not constitute a reduction in problem scope (as did the other reductions),
yet it made the overall problem computationally tractable.

This result will serve as a motivating factor for future research in performance-oriented composition of \ac{FSM} and model checking
problem specification based on traditional engineering models.

\subsection{Validation}
\label{sec:validation}

\subsubsection{Error Detection}
As described in \autoref{sec:assertion-generation}, we generate assertions on the model based
on component archetypes. For example, for breaker \texttt{CircuitBreakerEY162} we generate an assertion
that \texttt{CircuitBreakerEY162} will remain connected (a safety assertion), as shown in \autoref{lst:breaker-162-connected-assert}.

\begin{lstlisting}[language=NuSMV, label={lst:breaker-162-connected-assert}, caption={Breaker 162 Draw Assertion.}]
LTLSPEC G(CircuitBreakerEY166.state = connected)
\end{lstlisting}

The SLICED model was useful in fault determination. For example, it was discovered that sufficient power draw through \texttt{CircuitBreakerEY162} could
cause it to trip and become disconnected, as shown in the counter example in \autoref{lst:breaker-162-connected-trace}.

\begin{lstlisting}[language=NuSMV, label={lst:breaker-162-connected-trace}, caption={Counterexample to the Breaker Draw Assertion (subset of full NuSMV output).}]
-- specification  G CircuitBreakerEY162.state = connected  is false
-- as demonstrated by the following execution sequence
Trace Description: LTL Counterexample
Trace Type: Counterexample
-> State: 1.1 <-
CircuitBreakerEY162.state = connected
CircuitBreakerEY162.supplyingPower = TRUE
-> State: 1.2 <-
BankOne.draw = 11
CircuitBreakerEY162.draw = 11
-> State: 1.3 <-
CircuitBreakerEY162.state = broken
\end{lstlisting}

\paragraph*{Validation with Simulink}
The Virtual ADAPT model is designed to allow user \emph{injection} of error into the model.
\footnote{The ADAPT documentation uses the term ``fault'' however this paper uses definitions of ``fault'' and ``error'' from Avizienis et al.,
who define ``fault'' as the hypothesized cause of an error \cite{avizienis2004basic}.}
The objective of SLICED is to find errors in the design as specified (without injection of specific errors),
so to validate SLICED and Virtual ADAPT we manually introduced errors into the Virtual ADAPT model \emph{specification} (instead of injecting them at runtime).

\begin{figure}[t]
\centering{\includegraphics[width=1\linewidth]{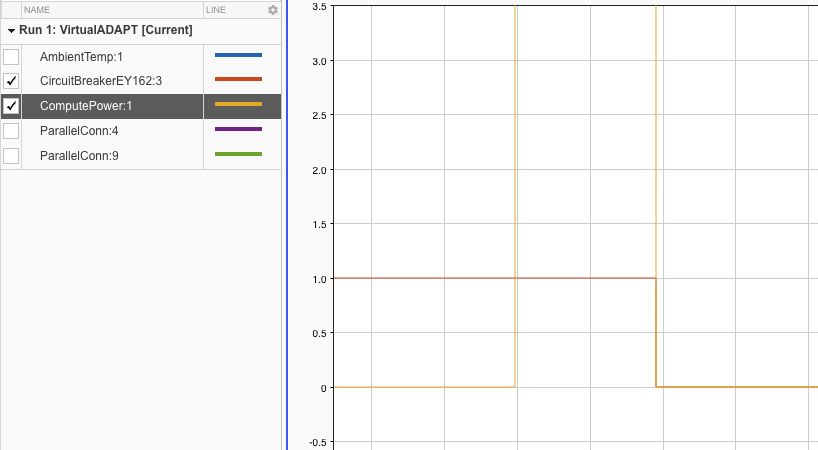}}
\caption{Example of simulation Virtual ADAPT demonstrating a high power power draw tripping CircuitBreakerEY162.}
\label{fig:virtualadapt-replication-one}
\end{figure}

We replicated the example detection of a circuit breaker tripping on excessive load by changing an actuator's parameters in Virtual ADAPT to increase its power consumption (as described by the counterexample generated by NuSMV).
\autoref{fig:virtualadapt-replication-one} shows a screenshot of Simulink's data viewer tracking the power draw from the
actuator (ComputerPower:1) and the state of the circuitbreaker (CircuitBreakerEY162:3). Note that the state of the circuit breaker changes
from connected (1) to disconnected (0) shortly after the increase in power consumption from the actuator.

\subsubsection{Fault Mitigation}

\paragraph*{Methodology}

To exercise SLICED as a mechanism for generating error correction plans with ADAPT,
we first used SLICED to generate a baseline ADAPT NuSMV specification, then removed some components to make it computationally tractable (as described in \autoref{sec:performance}).
Next, we manually modified that specificiation to set the initial state of a given component to an error state,
for example by initializing Battery2 as dead (shown in \autoref{lst:battery-nusmv-startdead}).

\begin{lstlisting}[language=NuSMV, label={lst:battery-nusmv-startdead}, caption={Battery module with dead initialization specified in NuSMV}]
MODULE BatteryStartDead(output1, output2, capacity)
VAR
  state : {nominal,dead,underRepair};
DEFINE
  supplyingPower := (state = nominal);
  draw := (output1.draw + output1.draw);
ASSIGN
  init(state) := dead;

  next(state) := case
    (draw > capacity) : dead;
    ((state = dead) & (draw = 0)) : underRepair;
    ((state = underRepair) & (draw = 0)) : nominal;
    TRUE : state;
  esac;
\end{lstlisting}

Next, we specified an assertion claiming that ``there is no way to return the system to a good state'' such
as the assertion in \autoref{lst:repair-battery-assertion}.

\begin{lstlisting}[language=NuSMV, label={lst:repair-battery-assertion}, caption={Repair assertion for ADAPT}]
-- NuSMV LTL assertion claiming that there is no future
-- state in which all of the listed
-- actuators are nominal and battery 1 is in a nominal state
-- with a power draw of two or less.
LTLSPEC G(
  NEMO.state != nominal |
  L1H.state != nominal |
  DCLoadBox.state != nominal |
  Load2G.state != nominal |
  Battery1.state != nominal |
  Battery1.draw > 2 )
\end{lstlisting}

Finally, we ran NuSMV on the modified model. If a repair plan was possible, NuSMV generated a sequence of steps that could
be taken by the user (recall that user-actionable operations are specified as non-deterministic in our model, so NuSMV can vary them at will to generate plan).

We determined that our use of NuSMV discovered repair plans that were valid in the context of the SLICED-generated \ac{FSM} (that is, they were legal sequences of transitions),
but that our generated \ac{FSM} did not facilitate constraints on the plan discovery so that other desired system constraints are maintained throughout execution of the plan.
This meant that we could generate viable repair plans for simple problems (e.g., turn a switch back on), but
that attempts to generate multi-step plans were often met with technically valid but practically infeasible results, such as rapidly switching relays on and off.

\section{Conclusions}
This paper described a methodology for generating \ac{FSM} that formally describe the behavior of components
in a system model. As a case study, this paper used the NASA Virtual \ac{ADAPT} Simulink model as a data source from which to
generate NuSMV specifications. We evaluated the performance of NuSMV on our generated \acp{FSM} and replicated findings from
analysis of the \ac{FSM} in Simulink.

\subsection{Results}
We described a process for creation of an \ac{FSM} by identifying component archetypes in the source model,
creating standardized state machines for each archetype, and assembling the final \ac{FSM} from component state machines.

\paragraph*{Performance}
We determined that naive generation of an \ac{FSM} from components in the Virtual ADAPT model yielded a state machine too large for
brute force exploration. We determined that further decomposition of the \ac{FSM} into partial models enabled exploration.

\paragraph*{Error Detection}
We described a process for generating assertions about components in a generated \ac{FSM}. We demonstrated
detection of a tripping circuit breaker, and replicated the trip in Simulink.

\paragraph*{Repair Plan Generation}
We determined that the level of fidelity and detail generated by SLICED is appropriate and viable for
generation of plans, but that the assertion language and counterexample
capability we evaluated with NuSMV were not ideal for plan generation because the \ac{FSM} we generated did not readily account
system constraints when generating a plan.

\paragraph*{Future Work}

The results of our experiment demonstrated that a formal model of abstract components derived from a high fidelity abstraction
can provide results that are replicable in the source conventional engineering model.
However, further work is required to verify our claim about the completeness of the SLICED methodology.
In particular, the methodology for translating Simulink blocks to NuSMV used in this experiment was based on heuristics. stronger
arguments about the correctness of the formal model will require rigorous translation.
Additional work should contribute toward improving the structure of generated models for subsystem separation
and on identifying solver optimization approaches to take advantage of the generated models.

Additional work is also needed to refine the generation of repair plans from conventional engineering models,
either by exploring new methods of problem specification, by using different analysis tools
(e.g., different model checkers or dedicated planning tools), or both.

\begin{acronym}[AIMFIRST]
\acro{AADL}{Architecture Analysis and Design Language}
\acro{ACVIP}{the Architecture-Centric Virtual Integration Process}
\acro{ADAPT}{Advanced Diagnostic and Prognostic Testbed}
\acro{AGREE}{Assume Guarantee Reasoning Environment}
\acro{BDD}{Binary Decision Diagram}
\acro{BMC}{Bounded Model Checking}
\acro{COI}{Cone of Influence}
\acro{CTL}{Computation Tree Logic}
\acro{DARPA}{Defense Advanced Research Projects Agency}
\acro{DFA}{Deterministic Finite Automata}
\acro{FACE}{Future Airborne Capability Environment}
\acro{FSM}{Finite State Machine}
\acro{GCD}{Greatest Common Denominator}
\acro{HACMS}{High-Assurance Cyber Military Systems}
\acro{HOST}{Hardware Open Systems Technologies}
\acro{LTL}{Linear Temporal Logic}
\acro{NFA}{Nondeterministic Finite Automata}
\acro{MBSE}{Model Based Systems Engineering}
\acro{SPICA}{Separation Platform for Integrating Complex Avionics}
\acro{SMV}{Symbolic Model Verification}
\acro{TLS}{Transport Layer Security}
\end{acronym}

%%
%% Bibliography
%%

%% Please use bibtex,
%

\bibliography{sliced-adapt-references}

\appendix

\end{document}